\documentclass[journal]{IEEEtran}

\usepackage{cite}
\usepackage{graphicx}
\usepackage{epstopdf}
\usepackage{amssymb, amsmath, amsthm}
\usepackage{amsfonts}
\usepackage{multirow}
\usepackage{mathrsfs}
\usepackage{color}
\usepackage{url}
\usepackage{flushend}
\usepackage{setspace}
\usepackage{placeins}
\usepackage{wrapfig}
\usepackage{booktabs, multicol, multirow}
\usepackage{slashbox}
\usepackage{makecell}
\usepackage[ruled]{algorithm2e}
\usepackage{soul}
\usepackage{diagbox}
\allowdisplaybreaks

\newtheorem{special cases}{Special Cases}

\newcommand{\ie}{\emph{i.e.}}
\newcommand{\eg}{\emph{e.g.}}

\newcommand{\tabincell}[2]{\begin{tabular}{@{}#1@{}}#2\end{tabular}}

\begin{document}

\bibliographystyle{IEEEtran}

\title{Detection of Information Hiding\\ at Anti-Copying 2D Barcodes}

\author{Ning~Xie,~\IEEEmembership{Senior Member,~IEEE}, Ji~Hu, Junjie~Chen, Qiqi~Zhang, and Changsheng~Chen,~\IEEEmembership{Member,~IEEE}
\vspace{-0.5cm}
\thanks{%
The authors are with the Guangdong Key Laboratory of Intelligent Information Processing, College of Information Engineering, Shenzhen University, Shenzhen, 518060, China (e-mail: ningxie@szu.edu.cn; cschen@szu.edu.cn).
}
}

\maketitle
\begin{abstract}
This paper concerns the problem of detecting the use of information hiding at anti-copying $2$D barcodes. Prior hidden information detection schemes are either heuristic-based or Machine Learning (ML) based.
The key limitation of prior heuristics-based schemes is that they do not answer the fundamental question of why the information hidden at a $2$D barcode can be detected.
The key limitation of prior ML-based information schemes is that they lack robustness because a printed 2D barcode is very much environmentally dependent, and thus an information hiding detection scheme trained in one environment often does not work well in another environment.
In this paper, we propose two hidden information detection schemes at the existing anti-copying $2$D barcodes.
The first scheme is to directly use the pixel distance to detect the use of an information hiding scheme in a $2$D barcode, referred as to the Pixel Distance Based Detection (PDBD) scheme.
The second scheme is first to calculate the variance of the raw signal and the covariance between the recovered signal and the raw signal, and then based on the variance results, detects the use of information hiding scheme in a $2$D barcode, referred as to the Pixel Variance Based Detection (PVBD) scheme. Moreover, we design advanced IC attacks to evaluate the security of two existing anti-copying $2$D barcodes.
We implemented our schemes and conducted extensive performance comparison between our schemes and prior schemes under different capturing devices, such as a scanner and a camera phone. Our experimental results show that the PVBD scheme can correctly detect the existence of the hidden information at both the 2LQR code and the LCAC $2$D barcode. Moreover, the probability of successfully attacking of our IC attacks achieves $0.6538$ for the 2LQR code and $1$ for the LCAC $2$D barcode.
\end{abstract}

\section{Introduction}
\subsection{Background and Motivation}
The Illegitimately-Copying (IC) attacks seriously hinder the application of Two-Dimensional (2D) barcodes as an anti-counterfeiting technique since a 2D barcode can be easily replicated with an off-the-shelf photocopier.
The IC attacks introduce large economic and reputational loss for an authorized manufacturer.
Some anti-copying 2D barcodes have been proposed to overcome the security risk of IC attacks but accompanying with some limitations, \eg, special printing materials or techniques, and physical unclonable function.
They have two limitations: high production cost and low universal applicability.
Recently, some new anti-copying 2D barcodes were proposed utilizing the additional distortion on the received 2D barcode introduced by IC attacks to effectively overcome the above limitations, such as Two-Level QR (2LQR) \cite{Tkachenko2016Two} code and Low-Cost Anti-Copying (LCAC) \cite{Xie2020Low} 2D barcode.
Specifically, the 2LQR code \cite{Tkachenko2016Two} and the LCAC 2D barcode \cite{Xie2020Low} embed subtle patterns and authentication message into the 2D barcode, respectively, to realize the anti-copying purpose.

This paper concerns the problem of detecting the use of information hiding at anti-copying 2D barcodes. Such information hiding detection schemes have the following applications.
First, it can be used to evaluate the security level of the existing anti-copying 2D barcodes.
For example, if the existence of the hidden information in a printed 2D barcode can be detected by an attacker, advanced IC attacks can be launched by the attacker.
Second, it can be used to detect illegal information that is hidden in a normal 2D barcode, \eg, Trojan virus or phishing websites.
Third, it can be used by military and law enforcement agencies to detect the leakage of confidential or even classified information.
Thus, the objective of this paper is to detect the existence of the hidden information in a printed 2D barcode and further to design advanced IC attacks to evaluate the security level of the existing anti-copying 2D barcodes.

\subsection{System Model}
In our system model, there are one sender and one receiver, where the sender prints a 2D barcode and sends it to the receiver two possible channels: a legal channel and an illegal channel.
The legal channel describes the Single Print and Capture (SPC) process whereas the illegal channel describes the Double Print and Capture (DPC) process.
The sender may or may not embed hidden information in the original message of a 2D barcode using the existing anti-copying 2D barcodes.
If the 2D barcode is indeed embedded with hidden information, the sender and the receiver often share a secret so that the sender can hide information with this secret and the receiver can decode the hidden information with the same secret.
We assume that our hidden information detector, which we call a monitor, captures the 2D barcode printed by the sender since a 2D barcode can be easily replicated with an off-the-shelf photocopier.
We assume that the monitor does not know a prior whether the 2D barcode carries hidden information or not, and if they do, we assume that the monitor does not know the shared secret between the sender and the receiver.
\subsection{Limitations of Prior Art}
Prior hidden information detection schemes are either heuristic-based or Machine Learning (ML) based \cite{Yadav2019Estimation}.
The key limitation of prior heuristics-based schemes is that they do not answer the fundamental question of why the information hidden at a 2D barcode can be detected.
The key limitation of prior ML-based information schemes is that they lack robustness because a printed 2D barcode is very much environmentally dependent, and thus an information hiding detection scheme trained in one environment often does not work well in another environment.
\subsection{Proposed Approach}
Our approach is based on the insight that embedding hidden information into a 2D barcode will inevitably have a negative impact on the decodability of the original message of a 2D barcode, such as the increase of the error probability at the receiver (as well as the monitor).
If the total errors introduced by the channel noise and the embedding operation exceeds the error-correcting capability of the original message of a 2D barcode at the receiver under an SPC process, then the presence of the hidden information becomes obvious and such anti-copying 2D barcode should not be applied in practice.
If the total errors introduced by the channel noise and the embedding operation does not exceed the error-correcting capability of the original message of a 2D barcode at the receiver, then the monitor can recover the original message. Based on the above insight, in our approach, after the monitor recovers the original message, it will re-encode and re-modulate the original message, and then compare the resulting signals, which we call recovered signals, with the raw signals that it received from the sender.
The key technical challenge is how to compare the recovered signals with the raw received signals so that we can detect the presence of the hidden information.
To address this challenge, we first propose our information detection approach at anti-copying 2D barcodes by comparing the recovered signals with the raw received signals.
Second, we derive a rigorous theoretical analysis of the test statistics of our approach for different hypotheses, such as the probability of a false alarm (PFA) and the probability of detection (PD).
Based on the theoretical analysis, we can calculate the optimal decision threshold using the Neyman-Pearson (NP) theorem by maximizing PD while ensuring that PFA does not exceed a threshold. Furthermore, we can measure the impact of existing anti-copying 2D barcodes or capturing devices, \eg, scanner or camera phone, on the detectability of our approach.
\subsection{Advantages over Prior Art}
We advance the state-of-the-art on detecting the information hidden at the existing anti-copying 2D barcodes from two fronts.
First, in comparison with prior heuristic-based schemes, our approach answers the fundamental question of why the information hidden at the existing anti-copying 2D barcodes can be detected.
Second, in comparison with prior ML-based schemes, our approach is much more robust because our approach is environmentally independent.
\subsection{Technical Challenges and Our Solution}
The first technical challenge is to construct a reasonable test statistic, as it directly determines the final detection performance of our approach.
To address this challenge, we propose two hidden information detection schemes at the existing anti-copying 2D barcodes.
The first scheme is to directly use the pixel distance to detect the use of an information hiding scheme in a 2D barcode, referred as to the Pixel Distance Based Detection (PDBD) scheme.
The second scheme is first to calculate the variance of the raw signal and the covariance between the recovered signal and the raw signal, and then based on the variance results, detects the use of information hiding scheme in a 2D barcode, referred as to the Pixel Variance Based Detection (PVBD) scheme.
The PVBD scheme has a better detection performance than the PDBD scheme because it magnifies the difference between the 2D barcode carrying no hidden information and the 2D barcode carrying hidden information.
In comparison, the PDBD scheme has two key advantages over the PVBD scheme.
First, the PDBD scheme runs faster than the PVBD scheme.
Second, the PDBD scheme can pinpoint the exact locations on the received $2$D barcode that are embedded with hidden information, whereas the PVBD scheme cannot.

The second technical challenge is to find the optimal test threshold for our detection schemes because either an SPC process or a DPC process is challenging to accurately be modeled and an inappropriate threshold inevitably introduces large false alarms or missed detections.
To address this challenge, we first present a simplified theoretical model for the test statistic of two proposed detection schemes using a generalized Gaussian distribution (GGD).
Then, based on the simplified theoretical, we obtain the optimal test thresholds of our detection schemes.

The third technical challenge is to effectively evaluate the security level of the existing anti-copying $2$D barcodes because there are some limitations in the existing IC attacks.
For example, a synthetic IC attack requires multiple versions of a legitimate $2$D barcode.
If a $2$D barcode generated by the merchant is unique, a monitor cannot launch an effective IC attack.
For another example, an ML-based IC attack requires sufficiently many trained samples to obtain a good attacking model, especially it is extremely challenging for a monitor to obtain the electronic version of a $2$D barcode containing hidden information.
To address this challenge, we design advanced IC attacks to evaluate the security of two existing anti-copying $2$D barcodes: the 2LQR code and the LCAC $2$D barcode.
\begin{enumerate}
  \item In the 2LQR code, the legitimate sender and receiver should share a pattern database in whole or in part for making an authentication decision.
      Thus, we propose two advanced IC attacks for the 2LQR code under two scenarios: Public Pattern Database (PPD) and Unknown Pattern Database (UPD).
      Under the PPD scenario, the goal of our IC attack is to first find the chosen patterns from the public pattern database and then based on the chosen patterns to launch an effective IC attack.
      Under the UPD scenario, our IC attack first estimates the size of a received pattern and then constructs an alternative pattern database through a binarization technique.
      At last, our IC attack finds the chosen patterns from the alternative pattern database and then based on the chosen patterns to launch an effective IC attack.
  \item For the LCAC $2$D barcode, based on the PDBD scheme, our IC attack should pinpoint the exact locations on the received $2$D barcode that are embedded with hidden information.
      Thus, we propose two advanced IC attacks for the LCAC $2$D barcode under two cases: All Constellation Points (ACP) and Separate Constellation Points (SCP).
      In the ACP case, the detection of embedded locations is designed on all constellation points, where we treat all constellation points equally.
      In the SCP case, the detection of embedded locations is separately designed on different constellation points, where we treat different constellation points individually.
\end{enumerate}

\subsection{Key Contributions and Results}
In this paper, we make three key contributions.
First, we propose two hidden information detection schemes at the existing anti-copying $2$D barcodes.
We explicitly analyze the PD and PFA of our schemes, derive their closed-form expressions, and obtain their optimal test thresholds.
Second, we design advanced IC attacks to evaluate the security of two existing anti-copying $2$D barcodes: the 2LQR code and the LCAC $2$D barcode.
In particular, our IC attacks can deal with the scenario of an unknown pattern database in the 2LQR code and can correctly pinpoint the embedded locations in the LCAC $2$D barcode.
Three, we implemented our schemes and conducted extensive performance comparison between our schemes and prior schemes under different capturing devices, such as a scanner and a camera phone.
Our experimental results show that the PVBD scheme can correctly detect the existence of the hidden information at both the 2LQR code and the LCAC $2$D barcode.
Moreover, the probability of successfully attacking of our IC attacks achieves $0.6538$ for the 2LQR code and $1$ for the LCAC $2$D barcode.


\section{Related Work}
%
%
\subsection{Anti-Copying $2$D barcodes}
\subsubsection{Special Printing Materials or Techniques}
These approaches exploit the special features of printing materials or printing techniques to defend against the IC attacks since the special features cannot be reproduced on purpose.
Adams et al. used specific printers to print a $2$D barcode for defending against IC attacks \cite{Adams2011A,Simske2010High}.
Marguerettaz et al. used a polymerized liquid crystal material with unique optical characteristics to print an anti-copying $2$D barcode \cite{Gremaud2015Identification}.
Wang et al. used a special halftone printing technology to print an anti-copying $2$D barcode that is invisible under visible light \cite{Wang2013Using}.
\subsubsection{Physical Unclonable Function (PUF)}
In the PUF, a stimulus is an input to a physical entity and we use the output as unique features.
The PUF is an unclonable response function since these features are determined by the internal physical structure, \eg, the certain unique texture of printing paper \cite{Voloshynovskiy2016Physical,Voloshynovskiy2012Towards}.
Recently, some researchers constructed the PUF for mobile imaging devices under a semi-controlled condition to extract microscopic textural features \cite{Wong2016Counterfeit,Wong2015A,Diephuis2013Physical}.

However, there are two limitations to the first two approaches: high production cost and low universal applicability, which hinders their promotion in extensive applications.
Actually, every P\&C process inevitably introduces additional distortion on the received $2$D barcode, which is an intrinsic channel feature of IC attacks.
Thus, the channel feature can be used to defend against IC attacks \cite{Claycomb2006Using,Picard2004Society,Goldman1987Non}.
Recently, based on the channel feature, some researchers proposed new low-cost $2$D barcodes to effectively overcome the limitations of the first two approaches, which are briefly introduced as follows.
\subsubsection{Two-Level QR ($2$LQR) Code}
Tkachenko et al. proposed the 2LQR code by replacing all black modules of a standard QR code with some black-and-white subtle patterns \cite{Tkachenko2016Two}.
These subtle patterns have two requirements: unknown to the third party and sensitive to a P\&C process.
The $2$LQR code has the following limitations.
First, the $2$LQR code introduces visually perceptual modification.
Second, the $2$LQR code requires a higher positioning accuracy of the capturing equipment or a higher proportion of the training sequence.
Thus, it is difficult to apply the $2$LQR code to the scenario that the camera of a mobile phone works as the capture device of the legitimate receiver.
Third, the $2$LQR code was designed for $2$D barcodes with two-order modulation, which cannot straightforwardly be extended to $2$D barcodes with higher-order modulation.
\subsubsection{Low-Cost Anti-Copying (LCAC) $2$D barcode}
Xie et al. proposed the LCAC $2$D barcode by exploiting the difference between the noise characteristics of legitimate and illegitimate channels \cite{Xie2020Low}.
Specifically, the sender of the LCAC $2$D barcode embeds an authentication message into the original message to realize the anti-copying purpose.
The LCAC $2$D barcode effectively overcomes the aforementioned limitations of the $2$LQR code.
\subsection{Existing IC Attacks}
%
%
\subsubsection{Direct IC Attack}
A monitor first captures a legitimate $2$D barcode.
Then, the monitor directly prints it to spoof the legitimate receiver.
\subsubsection{Synthetic IC Attack}
A monitor first collects multiple versions of a legitimate $2$D barcode.
Second, the monitor makes a preprocessing to launch an IC attack, \eg, a smoothing operation.
At last, the monitor prints the output of the preprocessing to spoof the legitimate receiver.
The basic idea of a synthetic IC attack is to suppress the distortion caused by a P\&C process through the preprocessing.
\subsubsection{Machine Learning (ML) Based IC Attack}
Yadav et al. proposed an efficient IC attack by using a supervised ML approach to launch an IC attack \cite{Yadav2019Estimation}.
This ML-based IC attack is realized through a neural network that consists of three parts: $5$ layers of encoders, $5$ layers of decoders, and one hidden layer that provides a useful compressed representation of the input \cite{Jorge2019A}.
First, a monitor collects sufficiently large numbers of legitimate $2$D barcodes and their corresponding electronic versions for training the neural network to obtain an attacking model.
Second, the monitor uses the attacking model to obtain the electronic version of an illegitimate $2$D barcode. At last, the monitor prints the output of the neural network to spoof the legitimate receiver.
The basic idea of an ML-based IC attack is to reconstruct the electronic version of a new $2$D barcode containing hidden information.

\section{System Model and Problem Statement}
\subsection{System Model}
\begin{figure}[!t]
\centering
\includegraphics[width=3.55in]{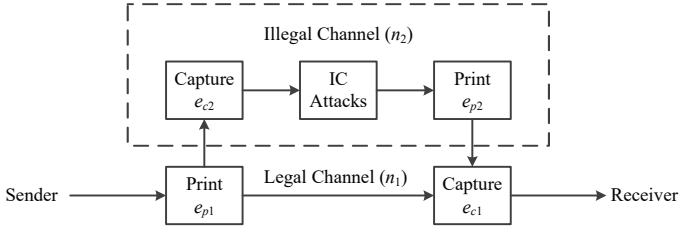}
\caption{System model of a $2$D barcode with two possible channels, \ie, a legitimate channel and an illegitimate channel.}
\label{Fig System_model_$2$D}
\vspace{-0.5cm}
\end{figure}
We illustrate the system model of a $2$D barcode with two possible channels in Fig. \ref{Fig System_model_$2$D}, where the lower branch represents a legitimate channel (an SPC process) and the dashed box represents an illegitimate channel (a DPC process).
Intuitively, the distortion and noise in a DPC process are more severe than those in an SPC process.
Specifically, we model the total noise in a legitimate channel as
\begin{equation}
{{e}_{1}}={{e}_{{{p}_{1}}}}\oplus {{e}_{c1}},
\label{equ1}
\end{equation}
where ${{e}_{{{p}_{1}}}}$ and ${{e}_{c1}}$ represent the noise components of the first printing process and the legitimate capture process, respectively, and '$\oplus $' represents the interaction of noise in different stages.
Then, we model the total noise in an illegitimate channel as
\begin{equation}
{{e}_{2}}={{e}_{{{p}_{1}}}}\oplus {{e}_{c1}}\oplus {{e}_{{{p}_{2}}}}\oplus {{e}_{c2}},
\label{equ2}
\end{equation}
where ${{e}_{c2}}$ and ${{e}_{{{p}_{2}}}}$ represent the noise components of the illegitimate capture process and the second printing process, respectively.
Based on the results of \cite{Xie2020Low,Zhang2019Accurate}, we can easily conclude that the variance of ${{e}_{2}}$ is significantly larger than that of ${{e}_{1}}$.

In the sender, an original message is generated, denoted as ${{I}_{o}}$.
In the meantime, hidden information is generated, denoted as ${{I}_{h}}$, \eg, subtle patterns for the $2$LQR code \cite{Tkachenko2016Two} or an authentication message for the LCAC $2$D barcode \cite{Xie2020Low}.
Note that the hidden information can be simultaneously generated at both the sender and the receiver according to the shared secret $K$, which is unknown to the third party.
Then, we embed the hidden information into the original message to obtain the electronic version of a legitimate $2$D barcode with $Q$-order modulation, denoted as $I$.
Specifically, in the $2$LQR code, all black modules of ${{I}_{o}}$ are replaced by the subtle patterns defined in ${{I}_{h}}$;
in the LCAC $2$D barcode, certain bits of the source message in ${{I}_{o}}$ are replaced by those of the authentication message defined in ${{I}_{h}}$.

Then, the sender prints $I$ and a receiver captures the printed $2$D barcode.
In other words, through a P\&C process, the receiver obtains a degraded version of $I$, denoted as $R$.
Through an equalizer to compensate for the channel distortion, the receiver obtains an equalization output ${{\hat{Y}}_{x}}$$\left( x=1,2 \right)$.
Here, $x=1$ represents an SPC process that ${{\hat{Y}}_{1}}$ is legitimate, whereas $x=2$ represents a DPC process that ${{\hat{Y}}_{2}}$ is illegitimate.

For defending against IC attacks, the receiver performs an authentication test by comparing ${{I}_{h}}$ with ${{\hat{Y}}_{x}}$.
From the mathematical point of view, we formulate the authentication test as a threshold test with hypotheses, given as follows.
\begin{equation}
\begin{matrix}
   {{\mathcal{H}}_{0}}: & d({{I}_{h}};{{{\hat{Y}}}_{x}})\le {{\theta }_{b}}  \\
   {{\mathcal{H}}_{1}}: & d({{I}_{h}};{{{\hat{Y}}}_{x}})>{{\theta }_{b}}  \\
\end{matrix}
,
\label{equ4}
\end{equation}
where ${{\mathcal{H}}_{0}}$ represents an SPC process that the received $2$D barcode is legitimate and ${{\mathcal{H}}_{1}}$ represents the opposite case.
Here, $d\left( x;y \right)$ is a comparison function that first extracts the features from $x$ and $y$, respectively, and then compares the two features to obtain a test statistic.
Specifically, in the $2$LQR code, $d\left( I;{{{\hat{Y}}}_{x}} \right)$ is defined as a Pearson correlation between the patterns of ${{I}_{h}}$ with the corresponding patterns of ${{\hat{Y}}_{x}}$ \cite{Tkachenko2016Two};
in the LCAC $2$D barcode, $d\left( I;{{{\hat{Y}}}_{x}} \right)$ is defined as the Bit Error Ratio (BER) between the authentication message in ${{I}_{h}}$ and that in ${{\hat{Y}}_{x}}$ \cite{Xie2020Low}.
Here, ${{\theta }_{b}}$ is a decision threshold that is determined according to the NP theorem.
Specifically, the optimal threshold is determined by making the PFA less than a predetermined upper bound, where the PFA is defined as accepting ${{\mathcal{H}}_{1}}$ when ${{\mathcal{H}}_{0}}$ is true.
\subsection{Problem Statements}
Besides the receiver, a monitor also can capture the $2$D barcode printed by the sender since a $2$D barcode can be easily replicated with an off-the-shelf photocopier.
%
%
If the existence of the hidden information in a printed $2$D barcode can be detected by other parties, \eg, a monitor or an attacker, the security of the existing anti-copying $2$D barcodes is challenged, \eg, advanced IC attacks can be launched by the monitor.
Thus, we will propose two hidden information detection schemes at the existing anti-copying $2$D barcodes in Section IV.
Moreover, there are some limitations in the existing IC attacks described in Section II.
\subsubsection{Direct  IC Attack}
Although this attack is the simplest, its attacking performance is poor.
\subsubsection{Synthetic IC Attack}
Although a synthetic IC attack has better-attacking performance than a direct IC attack, it has a stronger assumption, \ie, multiple versions of a legitimate $2$D barcode.
If a $2$D barcode generated by the merchant is unique, a monitor cannot launch a synthetic IC attack.
\subsubsection{ML-Based IC Attack}
Although an ML-based IC attack can achieve better-attacking performance than the other IC attacks, there are some limitations in \cite{Yadav2019Estimation}.
First, it requires sufficiently many trained samples to obtain a good attacking model, especially it is extremely challenging for a monitor to obtain the electronic version of a $2$D barcode containing hidden information.
Second, for different anti-copying $2$D barcodes, an ML-based IC attack should re-train a new attacking model.
Third, the approach in \cite{Yadav2019Estimation} was designed for $2$D barcodes with two-order modulation, which cannot straightforwardly be extended to $2$D barcodes with higher-order modulation

%
\section{Two Hidden Information Detection Schemes at Anti-Copying $2$D Barcodes}
\subsection{Description of Our Approach}
\begin{figure}[!t]
\centering
\includegraphics[width=3.55in]{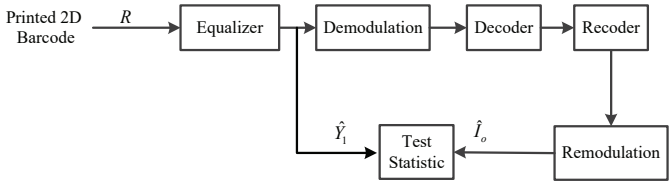}
\caption{Block diagram of our approach in the monitor.}
\label{Fig Diagram_of_monitor}
\vspace{-0.5cm}
\end{figure}
The block diagram of our approach in the monitor is illustrated in Fig. \ref{Fig Diagram_of_monitor}.
After obtained a printed $2$D barcode $R$ through an SPC process, the monitor first equalizes to compensate the channel distortion for a $2$D barcode with higher-order modulation ($Q>2$) by using the training symbols described in \cite{Xie2020Low,Zhang2019Accurate}.
Specifically, the equalizer block trains a fitting function to reflect the channel distortion by comparing the gray-scale values of the scanned training symbols with those of the considered ones.
When the fitting function is trained, an inverse fitting function is further established to correct the channel distortion. Through the equalizer block, the monitor obtains an equalization output ${{\hat{Y}}_{1}}$.
Then, through demodulation, decoder, recorder, and remodulation sequentially, the monitor obtains an estimated version of the original message, denoted as ${{\hat{I}}_{o}}$.
If the capability of correcting errors provided by the modulation and the channel coding together is larger than the total errors introduced by an SPC process, we assume that ${{\hat{I}}_{o}}={{I}_{o}}$.
This assumption holds in practical situations since the errors introduced by an SPC process is relatively small as compared with a DPC process. In other words, a reliable anti-copying $2$D barcode should provide an error-free bit sequence for the original message in an SPC process.

In our approach, the monitor performs a threshold test using the following hypotheses for detecting the presence of the hidden information, given as
\begin{equation}
\begin{matrix}
   {{\mathcal{H}}_{0}}: & {{{\hat{Y}}}_{1}}\text{ without containing }{{I}_{h}}  \\
   {{\mathcal{H}}_{1}}: & {{{\hat{Y}}}_{1}}\text{ with containing }{{I}_{h}}  \\
\end{matrix}
.
\label{equ5}
\end{equation}
Thus, the equalization output under two hypotheses can be respectively expressed as
\begin{equation}
{{\hat{Y}}_{1}}\left| _{{{\mathcal{H}}_{0}}} \right.={{I}_{o}}\oplus {{e}_{1}},
\label{equ6}
\end{equation}
\begin{equation}
{{\hat{Y}}_{1}}\left| _{{{\mathcal{H}}_{1}}} \right.=I\oplus {{e}_{1}}.
\label{equ7}
\end{equation}

Based on \eqref{equ5}, we propose two hidden information detection schemes at anti-copying $2$D barcodes by comparing the gray value of each pixel in ${{\hat{Y}}_{\text{1}}}$ and that in ${{\hat{I}}_{o}}$.
For example, the monitor constructs a test statistic $\delta $ and then makes a detection decision ${{\varphi }_{m}}$ on $\delta $ according to
\begin{equation}
{{\varphi }_{m}}=\left\{ \begin{matrix}
   0, & \delta <{{\theta }_{m}}  \\
   1, & \delta \ge {{\theta }_{m}}  \\
\end{matrix} \right.
,
\label{equ8}
\end{equation}
where ${{\theta }_{m}}$ is the test threshold using the hypotheses defined in \eqref{equ5}.
Here, ${{\varphi }_{m}}=1$ represents the hypothesis ${{\mathcal{H}}_{1}}$ in \eqref{equ5} while ${{\varphi }_{m}}=0$ represents the opposite hypothesis.
The optimal value of ${{\theta }_{m}}$ is determined by ${{\varepsilon }_{\text{PFA}}}$ according to the NP theorem, where ${{\varepsilon }_{\text{PFA}}}$ is the upper bound of the PFA allowed by the monitor.
\subsubsection{Pixel Distance Based Detection (PDBD) Scheme}
The first scheme is to directly use the pixel distance to detect the use of an information hiding scheme in a $2$D barcode, referred as to the Pixel Distance Based Detection (PDBD) scheme.
In the PDBD scheme, the test statistic is denoted as
\begin{equation}
\delta =\sum\limits_{i=1}^{{{N}_{d}}}{\left| {{{\hat{I}}}_{o}}\left( i \right)-{{{\hat{Y}}}_{1}}\left( i \right) \right|},
\label{equ9}
\end{equation}
where ${{N}_{d}}$ is the total number of pixels in a received $2$D barcode. Here, ${{\hat{I}}_{o}}\left( i \right)$ and ${{\hat{Y}}_{1}}\left( i \right)$ represent the $i\text{th}$ gray value of each pixel in ${{\hat{I}}_{o}}$ and ${{\hat{Y}}_{\text{1}}}$, respectively.

For different hypotheses defined in \eqref{equ5}, the test statistic in \eqref{equ9} can be respectively expressed by
\begin{eqnarray}
\begin{aligned}
{{\left. \delta  \right|}_{{{\mathcal{H}}_{0}}}}=\sum\limits_{i=1}^{{{N}_{d}}}{\left| {{I}_{o}}\left( i \right)-\left( {{I}_{o}}\left( i \right)\oplus {{e}_{1}}\left( i \right) \right) \right|} \\
\end{aligned}
,
\label{equ10}
\end{eqnarray}
\begin{eqnarray}
\begin{aligned}
{{\left. \delta  \right|}_{{{\mathcal{H}}_{1}}}}=\sum\limits_{i=1}^{{{N}_{d}}}{\left| {{I}_{o}}\left( i \right)-\left( I\left( i \right)\oplus {{e}_{1}}\left( i \right) \right) \right|} \\
\end{aligned}
,
\label{equ11}
\end{eqnarray}
where ${{I}_{o}}\left( i \right)$, $I\left( i \right)$, and ${{e}_{1}}\left( i \right)$ represent the $i\text{th}$ gray value of each pixel in ${{I}_{o}}$, $I$, and ${{e}_{1}}$, respectively.
\subsubsection{Pixel Variance Based Detection (PVBD) Scheme}
The second scheme is first to calculate the variance of ${{\hat{Y}}_{\text{1}}}$ and the covariance between  ${{\hat{I}}_{o}}$ and ${{\hat{Y}}_{\text{1}}}$, and then based on the variance results, detects the use of an information hiding scheme in a $2$D barcode, referred as to the Pixel Variance Based Detection (PVBD) scheme.
In the PVBD scheme, the test statistic is denoted as
\begin{equation}
\delta =1-\frac{\operatorname{cov}\left( {{{\hat{I}}}_{o}},{{{\hat{Y}}}_{1}} \right)}{\operatorname{var}\left( {{{\hat{Y}}}_{1}} \right)},
\label{equ12}
\end{equation}
where $\operatorname{var}\left( {{{\hat{Y}}}_{1}} \right)$ and $\operatorname{cov}\left( {{{\hat{I}}}_{o}},{{{\hat{Y}}}_{1}} \right)$ represent the sample variance of ${{\hat{Y}}_{\text{1}}}$ and the sample covariance between ${{\hat{I}}_{o}}$ and ${{\hat{Y}}_{\text{1}}}$, expressed as
\begin{equation}
\operatorname{var}\left( {{{\hat{Y}}}_{1}} \right)=\frac{1}{{{N}_{d}}}\sum\limits_{i=1}^{{{N}_{d}}}{{{\left( {{{\hat{Y}}}_{1}}\left( i \right)-\frac{1}{{{N}_{d}}}\sum\limits_{i=1}^{{{N}_{d}}}{{{{\hat{Y}}}_{1}}\left( i \right)} \right)}^{2}}},
\label{equ13}
\end{equation}
\begin{eqnarray}
\begin{aligned}
   \operatorname{cov}\left( {{{\hat{I}}}_{o}},{{{\hat{Y}}}_{1}} \right)&=\frac{1}{{{N}_{d}}}\sum\limits_{i=1}^{{{N}_{d}}}{\left( {{{\hat{I}}}_{0}}\left( i \right){{{\hat{Y}}}_{1}}\left( i \right) \right)}\\ &-\frac{1}{{{N}_{d}}}\sum\limits_{i=1}^{{{N}_{d}}}{{{{\hat{Y}}}_{1}}\left( i \right)}-\frac{1}{{{N}_{d}}}\sum\limits_{i=1}^{{{N}_{d}}}{{{{\hat{I}}}_{0}}\left( i \right)} \\
\end{aligned}
.
\label{equ14}
\end{eqnarray}
The basic idea of the PVBD scheme is to utilize the fact that the variance of ${{\hat{Y}}_{\text{1}}}$ under different hypotheses has a significant difference.
For different hypotheses defined in \eqref{equ5}, the test statistic in \eqref{equ12} can be respectively expressed by
\begin{equation}
{{\left. \delta  \right|}_{{{\mathcal{H}}_{0}}}}=1-\frac{\operatorname{cov}\left( {{I}_{o}},{{I}_{o}}\oplus {{e}_{1}} \right)}{\operatorname{var}\left( {{I}_{o}}\oplus {{e}_{1}} \right)},
\label{equ15}
\end{equation}
\begin{equation}
{{\left. \delta  \right|}_{{{\mathcal{H}}_{1}}}}=1-\frac{\operatorname{cov}\left( {{I}_{o}},I\oplus {{e}_{1}} \right)}{\operatorname{var}\left( I\oplus {{e}_{1}} \right)}.
\label{equ16}
\end{equation}

The PVBD scheme has a better detection performance than the PDBD scheme because it magnifies the difference between ${{I}_{o}}$ and $I$.
In comparison, the PDBD scheme has two key advantages over the PVBD scheme.
First, the PDBD scheme runs faster than the PVBD scheme.
Second, the PDBD scheme can pinpoint the exact locations on the received $2$D barcode that are embedded with hidden information, whereas the PVBD scheme cannot, which will be verified in the next section.
In the next subsection, we present a simplified theoretical model for the test statistic of two proposed schemes for obtaining their optimal test thresholds.
In the last subsection, we will present experimental results to verify the theoretical analysis of our approach.
\subsection{Performance Analysis of Our Approach}
Through observing the experimental results, we find that we can use a generalized Gaussian distribution (GGD) \cite{Younan2010Moments} to well describe the behavior of the test statistics of our approach, which will be verified in Section VI.B.
A GGD random variable (RV) is denoted as $x\sim \mathcal{G}\mathcal{G}\mathcal{D}\left( \mu ,{{\sigma }^{2}},\gamma  \right)$, where $\mu $ is the mean, ${{\sigma }^{2}}$ is the variance, and $\gamma $ is the shape factor.
According to \cite{Wang2006A}, we respectively express the Probability Distribution Function (PDF) and the cumulative distribution function (CDF) of $x$ as
\begin{equation}
{{f}_{X}}\left( x \right)=\frac{\gamma \eta \left( \sigma ,\gamma  \right)}{2\Gamma \left( 1/\gamma  \right)}\exp \big[ -{{\left( \eta \left( \sigma ,\gamma  \right)\left| x-\mu  \right| \right)}^{\gamma }} \big],
\label{equ17}
\end{equation}
\begin{equation}
F_X (x) = \frac{1}{2}+ \mbox{sgn}(x-\mu)\frac{\kappa \big[1/\gamma,(\left| x-\mu \right| \eta(\sigma, \gamma))\gamma \big]}{2\Gamma(1/\gamma)},
\label{equ18}
\end{equation}
where $\eta \left( \sigma ,\gamma  \right)=\frac{1}{\sigma }\sqrt{\frac{\Gamma \left( 3/\gamma  \right)}{\Gamma \left( 1/\gamma  \right)}}$, $\kappa \left( \cdot  \right)$ is the lower incomplete gamma function, $\Gamma \left( \cdot  \right)$ is the gamma function, and $\text{sgn}\left( x \right)$ represents a symbol decision function, \ie, $\text{sgn}\left( x \right)=1$, if $x\ge 0$, and $\text{sgn}\left( x \right)=-1$ otherwise.

According to \cite{Xie2020Low,Baierlein2004Probability}, we can estimate three parameters of a GGD distribution for the test statistic of our approach from experimental results.
We assume that there are ${{M}_{s}}$ samples of $\delta \left( j \right)$, $j=1,\ldots ,{{M}_{s}}$ for different hypotheses to estimate parameters of a GGD distribution. First, the sample mean and the sample variance are respectively obtained as
\begin{equation}
\mu \left( \delta  \right)=\frac{1}{{{M}_{s}}}\underset{j=1}{\overset{{{M}_{s}}}{\mathop \sum }}\,\delta \left( j \right),
\label{equ19}
\end{equation}
\begin{equation}
{{\sigma }^{2}}\left( \delta  \right)=\frac{1}{{{M}_{s}}}\sum\limits_{j=1}^{{{M}_{s}}}{{{\left( \delta \left( j \right)-\mu \left( \delta  \right) \right)}^{2}}}.
\label{equ20}
\end{equation}
Second, according to the results of \cite{Wang2006A,Sharifi1995Estimation} , we construct a generalized Gaussian ratio function $r\left( \gamma \left( \delta  \right) \right)$, defined as
\begin{equation}
r\left( \gamma \left( \delta  \right) \right)=\frac{{{\sigma }^{2}}\left( \delta  \right)}{{{\left( \frac{1}{{{M}_{s}}}\underset{j=1}{\overset{{{M}_{s}}}{\mathop \sum }}\,\left| \delta \left( j \right)-\mu \left( \delta  \right) \right| \right)}^{2}}}=\rho,
\label{equ21}
\end{equation}
where
\begin{equation}
\rho =\frac{\Gamma \left( 1/\gamma \left( \delta  \right) \right)\Gamma \left( 3/\gamma \left( \delta  \right) \right)}{{{\Gamma }^{2}}\left( 2/\gamma \left( \delta  \right) \right)}.
\label{equ22}
\end{equation}
Then, a feasible solution of $\gamma \left( \delta  \right)$ can be found as,
\begin{equation}
\gamma \left( \delta  \right)={{r}^{-1}}\left( \rho  \right),
\label{equ23}
\end{equation}
where we use an exhausted search approach for solving \eqref{equ23} to obtain an estimate of $\gamma \left( \delta  \right)$.
We calculate the PFA of our approach as
\begin{equation}
{{P}_{\text{FA}}}=\Pr \left\{ {{\left. \delta  \right|}_{{{\mathcal{H}}_{0}}}}>{{\theta }_{m}} \right\}=1-{{F}_{X}}({{\theta }_{m}}).
\label{equ24}
\end{equation}
By setting ${{P}_{\text{FA}}}\le {{\varepsilon }_{\text{PFA}}}$, we obtain the optimal value of ${{\theta }_{m}}$ as
\begin{equation}
\theta _{m}^{0}=\frac{{{\kappa }^{-1}}\left[ \frac{\text{1}}{\gamma \left( \delta  \right)},\left( 1-2{{\varepsilon }_{\text{PFA}}} \right)\Gamma \left( \frac{\text{1}}{\gamma \left( \delta  \right)} \right) \right]}{\eta \left( \sigma \left( \delta  \right),\gamma \left( \delta  \right) \right)\gamma \left( \delta  \right)}\text{+}\mu \left( \delta  \right),
\label{equ25}
\end{equation}
where ${{\kappa }^{-1}}\left( \cdot  \right)$ is the inverse of the incomplete lower gamma function.
At last, we calculate the Probability of Detection (PD) of our approach as
\begin{equation}
{{P}_{D}}=\Pr \left\{ {{\left. \delta  \right|}_{{{\mathcal{H}}_{1}}}}>\theta _{m}^{0} \right\}\text{=}1-{{F}_{X}}(\theta _{m}^{0}).
\label{equ26}
\end{equation}
\subsection{Experimental Results of Our Approach}
In this subsection, we compare the experimental results of our proposed detection schemes with their theoretical results, where two hypotheses defined in \eqref{equ5} are considered.
In our experiment, we put the same $2$D barcodes on an A4 paper 20 times and print it with a printer, and then scan each $2$D barcode with a scanner 50 times or with a camera phone 50 times.
Thus, each $2$D barcode is captured by 1000 times.
We set ${{\varepsilon }_{\text{PFA}}}\text{=}0.01$.
The experimental settings are given as follows:
\begin{itemize}
\item
Printer: HP LaserJet P$1108$ with $1200$ DPI;
\item
Printing Material: A$4$ paper with weight $120$ g/m$^2$ from the Xerox;
\item
Scanner: BENQ K$810$ scanner in $1200$ DPIE;
\item
Camera Phone: HONOR V$20$ with $4800$MP pixels;
\item
Capture Angle: Within $10$ degrees between the barcode image plane and the camera sensor plane;
\item
Capture Distance: About $15$ cm in the in-focus case;
\item
Lighting: $300$-$350$ lux for the bright case and $100$-$150$ lux for the dim case.
\end{itemize}

Based on the description in \cite{Tkachenko2016Two} and \cite{Xie2020Low}, the parameters of two existing anti-copying $2$D barcodes are respectively given as follows:
\begin{itemize}
\item
$2$LQR Code Design: A barcode with $25\times 25$ modules with actual size $1.2\times 1.2$ cm$^2$, $b=42\%$, and ${{L}_{b}}\text{=}225$;
Since the size of each pattern is $12\times 12$ pixels, ${{N}_{d}}\text{=}25\times 25\times 12\times 12\text{=90000}$ pixels;
\item
LCAC $2$D Barcode Design: A multilevel $2$D barcode with $47\times 47$ modules with actual size $3.2\times 3.2$ cm$^2$; The modulation order is 4, \ie, $Q=4$ and the constellation points are set as $\left\{ 40,100,160,220 \right\}$. Since the size of each module is $32\times 32$ pixels,   ${{L}_{d}}\text{=}2209$, ${{N}_{d}}\text{=}2209\times 32\times 32\text{=2262016}$ pixels.
\end{itemize}

\begin{figure}[!t]
\centering
\includegraphics[width=3.55in]{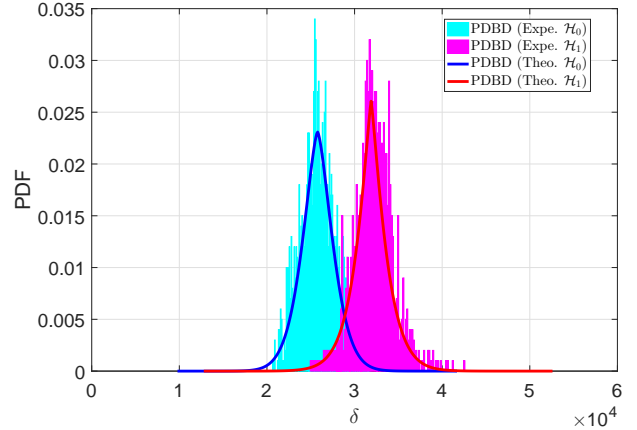}
\caption{Comparison of histograms and theoretical results about the test statistic of the PDBD scheme on the LCAC $2$D barcode, where a camera phone is used as the capturing device.}
\label{Fig PDBD}
\vspace{-0.5cm}
\end{figure}
\begin{figure}[!t]
\centering
\includegraphics[width=3.55in]{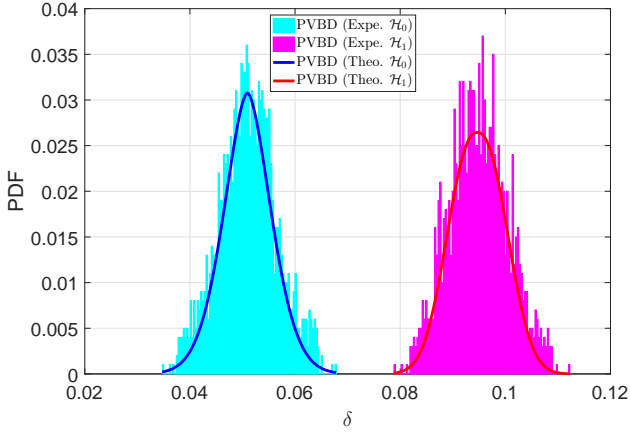}
\caption{Comparison of histograms and theoretical results about the test statistic of the PVBD scheme on the LCAC $2$D barcode, where a camera phone is used as the capturing device.}
\label{Fig PVBD}
\vspace{-0.5cm}
\end{figure}

We first present the distribution of the test statistic of our two schemes for the LCAC $2$D barcode, as illustrated in Fig. \ref{Fig PDBD} and Fig. \ref{Fig PVBD}, respectively, where both histograms and theoretical results are presented.
From Fig. \ref{Fig PDBD} and Fig. \ref{Fig PVBD}, we observe that the experimental results match well with GGD approximations. By comparing the results of Fig. \ref{Fig PDBD} and those of Fig. \ref{Fig PVBD}, we observe that the PVBD scheme has much better detection performance than the PDBD scheme since the distribution distance between two hypotheses in the PVBD scheme is significantly larger than that in the PDBD scheme.
This is because the operation divided by the variance of different hypotheses in \eqref{equ12} magnifies the difference between the $2$D barcode carrying no hidden information ${{I}_{o}}$ and the $2$D barcode carrying hidden information $I$, which verifies the aforementioned conclusion.
Moreover, since the $2$LQR code introduces visually perceptual modification, we do not present the histograms and theoretical results of our schemes on the $2$LQR code to save the page space.
In other words, the existence of the hidden information in the $2$LQR code is even visually detectable.

\begin{table}
\centering
\caption{Comparison of theoretical results, simulation results and experimental results of the PD of the PDBD scheme under different capturing devices.}
\footnotesize
\begin{tabular}{|c|c|c|c|c|}
  \hline
  \tabincell{c}{Anti-Copying\\$2$D Barcode} & \multicolumn{2}{c|}{$2$LQR}  & \multicolumn{2}{c|}{LCAC} \\
  \hline
  \tabincell{c}{Capturing\\Device} & Scanner & Camera Phone & Scanner & Camera Phone \\
  \hline
  Theoretical & $1$ & $1$ & $0.67$ &  $0.68$ \\
  \hline
  Simulation & $1$ & $1$ & $0.67$ &  $0.68$ \\
  \hline
  Experimental & $1$ & $1$ & $0.63$ &  $0.65$ \\
  \hline
\end{tabular}
\label{Table PDBD}
\vspace{0.2cm}
\end{table}

\begin{table}
\centering
\caption{Comparison of theoretical results, simulation results and experimental results of the PD of the PVBD scheme under different capturing devices.}
\footnotesize
\begin{tabular}{|c|c|c|c|c|}
  \hline
  \tabincell{c}{Anti-Copying\\$2$D Barcode} & \multicolumn{2}{c|}{$2$LQR}  & \multicolumn{2}{c|}{LCAC} \\
  \hline
  \tabincell{c}{Capturing\\Device} & Scanner & Camera Phone & Scanner & Camera Phone \\
  \hline
  Theoretical & $1$ & $1$ & $1$ &  $1$ \\
  \hline
  Simulation & $1$ & $1$ & $1$ &  $1$ \\
  \hline
  Experimental & $1$ & $1$ & $1$ &  $1$ \\
  \hline
\end{tabular}
\label{Table PVBD}
\vspace{0.2cm}
\end{table}

For evaluating the accuracy of our theoretical analysis, we compare the theoretical results, simulation results and experimental results of the PD of our two schemes under different capturing devices, as presented in Tab. \ref{Table PDBD} and Tab. \ref{Table PVBD}, respectively, where we set ${{\varepsilon }_{\text{PFA}}}\text{=}0.01$.
Here, the theoretical results are calculated through \eqref{equ26}.
For obtaining the simulation results, we first obtain the parameters of a GGD through \eqref{equ19}, \eqref{equ20}, and \eqref{equ23};
Second, based on these parameters, we generate a GGD sequence to simulate an SPC process;
Third, through the simulated SPC process, we obtain different equalization outputs under two hypotheses;
At last, we obtain the simulation results through \eqref{equ8}.
From Tab. \ref{Table PDBD} and Tab. \ref{Table PVBD}, we observe the following conclusions.
First, the theoretical results perfectly match the simulation results as we expected while the theoretical results very close to the experimental results;
Second, the existence of the hidden information in the $2$LQR code is easier detected than that in the LCAC $2$D barcode;
Third, the PVBD scheme has much better detection performance than the PDBD scheme;
At last, the detection performance under the camera phone is better than that under the scanner since the camera phone has better capturing resolution than the scanner in our experiment setup.
\section{Advanced IC Attacks}
%
%
\subsection{Our IC Attack for the $2$LQR Code}
To facilitate the introduction of our IC attack for the $2$LQR code, we first introduce some notations about the $2$LQR code.
In the $2$LQR code, each black-and-white subtle pattern consists of ${{N}_{p}}\times {{N}_{p}}$ pixels and has the same black pixel density, denoted by $b$.
Specifically, if there are ${{N}_{k}}$ black pixels in a subtle pattern, the value of $b$ is calculated by $b\text{=}{{N}_{k}}/N_{p}^{2}$.
We denote the $j\text{th}$ patterns in ${{I}_{h}}$ and ${{\hat{Y}}_{1}}$ as ${{P}_{j}}$ and ${{S}_{j}}$, respectively, where $j=1,\ldots ,{{L}_{b}}$ and ${{L}_{b}}$ is the total number of black modules in a $2$LQR code.
We calculate the Pearson correlation between ${{P}_{j}}$ and ${{S}_{j}}$ as
\begin{eqnarray}
\begin{aligned}
&\text{pcor}({{P}_{j}},{{S}_{j}}) = \\ &\frac{\sum\limits_{w=1}^{{{N}_{p}}}{\sum\limits_{h=1}^{{{N}_{p}}}{{{{\bar{P}}}_{j}}(w,h){{{\bar{S}}}_{j}}(w,h)}}}{\sqrt{\sum\limits_{w=1}^{{{N}_{p}}}{\sum\limits_{h=1}^{{{N}_{p}}}{{{\left( {{{\bar{P}}}_{j}}(w,h) \right)}^{2}}}}}\sqrt{\sum\limits_{w=1}^{{{N}_{p}}}{\sum\limits_{h=1}^{{{N}_{p}}}{{{\left( {{{\bar{S}}}_{j}}(w,h) \right)}^{2}}}}}} \\
\end{aligned}
,
\label{equ27}
\end{eqnarray}

Here, we denote ${{\bar{P}}_{j}}(w,h)$ and ${{\bar{S}}_{j}}(w,h)$ as follows, respectively,
\begin{equation}
{{\bar{P}}_{j}}(w,h)={{P}_{j}}(w,h)-\frac{1}{N_{p}^{2}}\sum\limits_{w=1}^{{{N}_{p}}}{\sum\limits_{h=1}^{{{N}_{p}}}{{{P}_{j}}(w,h)}},
\label{equ28}
\end{equation}
\begin{equation}
{{\bar{S}}_{j}}(w,h)={{S}_{j}}(w,h)-\frac{1}{N_{p}^{2}}\sum\limits_{w=1}^{{{N}_{p}}}{\sum\limits_{h=1}^{{{N}_{p}}}{{{S}_{j}}(w,h)}},
\label{equ29}
\end{equation}
where ${{P}_{j}}(w,h)$ and ${{S}_{j}}(w,h)$ are the gray value at the $w\text{th}$ row and the $h\text{th}$ column of  ${{P}_{j}}$ and ${{S}_{j}}$, respectively.

In the $2$LQR code, the legitimate sender and receiver should share a pattern database in whole or in part for making an authentication decision.
Thus, we propose two advanced IC attacks for the $2$LQR code under two scenarios: Public Pattern Database (PPD) and Unknown Pattern Database (UPD).
The PPD scenario is possible in a practical situation because a monitor can purchase different $2$LQR codes multiple times to recover the pattern database.
Under the PPD scenario, the goal of our IC attack is to first find the chosen patterns from the public pattern database and then based on the chosen patterns to launch an effective IC attack.
Then, under the UPD scenario, the design of our IC attack becomes more challenging as compared with the PPD scenario since the pattern database is unknown and a monitor has only one $2$LQR code.
Under the UPD scenario, our IC attack first estimates the size of a received pattern and then constructs an alternative pattern database through a binarization technique.
At last, our IC attack finds the chosen patterns from the alternative pattern database and then based on the chosen patterns to launch an effective IC attack.
\subsubsection{PPD Scenario}
For comparing the difference between the public pattern database and the alternative pattern database, based on the description in \cite{Tkachenko2016Two}, we first introduce the characteristics of the patterns in the public pattern database.
The patterns in the public pattern database have the following characteristics:
\begin{itemize}
\item
They have the pixels with the same size;
\item
They are binary;
\item
They have the same black pixel density, \ie, the value of $b$ keeps the same for each pattern;
\item
They have spectra related to them.
\end{itemize}

\begin{figure}[!t]
\centering
\includegraphics[width=1.55in]{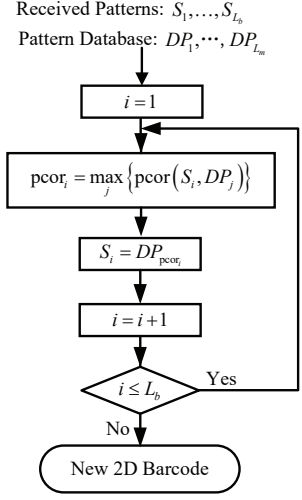}
\caption{Flow chart of our IC attack for the $2$LQR code.}
\label{Fig Flow_chart_$2$LQR}
\vspace{-0.5cm}
\end{figure}
We assume ${{L}_{b}}$ received patterns, denoted as ${{S}_{i}}$ ($i=1,\ldots ,{{L}_{b}}$), and ${{L}_{m}}$ patterns in the public pattern database, denoted as $D{{P}_{j}}$ ($j=1,\ldots ,{{L}_{m}}$).
We illustrate the flow chart of our IC attack for the $2$LQR code in Fig. \ref{Fig Flow_chart_$2$LQR}.
We introduce the detailed processes of our approach in \textbf{Algorithm 1}.

\begin{algorithm}
\caption{Our IC attack for the $2$LQR code}
\KwIn{${{S}_{i}}$ ($i=1,\ldots ,{{L}_{b}}$);$D{{P}_{j}}$ ($j=1,\ldots ,{{L}_{m}}$)}
\KwOut{new $2$D barcode}
  \For{$i=1;i\le {{L}_{b}}$}
  {
    $\text{pco}{{\text{r}}_{i}}=\underset{j}{\mathop{\max }}\,\left\{ \text{pcor}\left( {{S}_{i}},D{{P}_{j}} \right) \right\}$\\
    Replace ${{S}_{i}}$ by the chosen pattern $D{{P}_{\text{pco}{{\text{r}}_{i}}}}$
  }
  return new $2$D barcode;
\end{algorithm}
\subsubsection{UPD Scenario}
In this scenario, although the pattern database is unknown and only one $2$LQR code is available, the size of a received pattern is relatively easily estimated.
For example, after observed multiple received patterns with a high-resolution scanner, we look for some pixels with strict square shapes.
Based on these chosen pixels, we can estimate the size of a pixel and then can calculate the size of a received pattern, \ie, the ratio between the length of the received pattern to the length of a pixel.
Through multiple experiments based on the aforementioned public pattern database, the estimation accuracy on the size of a received pattern is perfect.
However, if we straightforwardly construct an entire pattern database only based on the size of a received pattern, the dimensionality of the entire pattern database becomes huge, \ie, ${{2}^{N_{p}^{2}}}$.
The dimensionality of the entire pattern database inevitably introduces prohibitive requirements on the storage space and the computational complexity, which hinders the application of our approach in practice.
Thus, it is a wise option to construct an alternative pattern database, whose dimensionality is significantly smaller than that of the entire pattern database.
\begin{figure}[!t]
\centering
\includegraphics[width=1.55in]{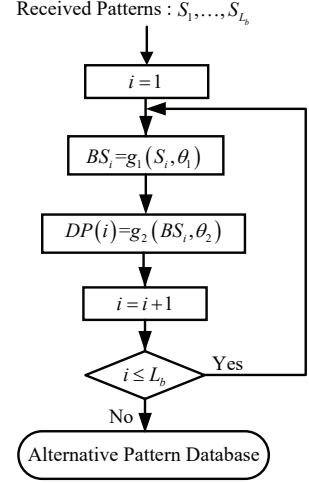}
\caption{Flow chart of constructing an alternative pattern database for the $2$LQR code.}
\label{Fig Flow_chart_database}
\vspace{-0.5cm}
\end{figure}
%

We illustrate the flow chart of constructing an alternative pattern database for the $2$LQR code in Fig. \ref{Fig Flow_chart_database}.
Here, we denote two functions. First, ${{g}_{1}}\left( {{S}_{i}},{{\theta }_{1}} \right)$ represents a binarization function that binarizes the received pattern ${{S}_{i}}$ with a threshold ${{\theta }_{1}}$ to obtain a binary pattern $B{{S}_{i}}$.
%
%
Second, ${{g}_{2}}\left( B{{S}_{i}},{{\theta }_{2}} \right)$ is to construct a sub-database of the alternative pattern database $DP\left( i \right)$ ($i=1,\ldots ,{{L}_{b}}$) based on a binary pattern $B{{S}_{i}}$ and a threshold ${{\theta }_{2}}$, where the size of the previous ${{L}_{b}}-1$ sub-databases is denoted by $\left\lfloor {{L}_{m}}/{{L}_{b}} \right\rfloor $, and that of the last sub-database patter is ${{L}_{m}}-\left( {{L}_{b}}-1 \right)\left\lfloor {{L}_{m}}/{{L}_{b}} \right\rfloor $, and $\left\lfloor \cdot  \right\rfloor $ is a floor function.
Specifically, we randomly modify the gray values of some pixels in $B{{S}_{i}}$ to obtain a new binary pattern, \eg, black pixels replaced by white pixels and vice versa.
Then we calculate the Pearson correlation between the new binary pattern and $B{{S}_{i}}$.
If the value of the Pearson correlation is no less than ${{\theta }_{2}}$ and the new binary pattern does not exist in the sub-database $DP\left( i \right)$, we put the new binary pattern into $DP\left( i \right)$.
The above processes continue until $DP\left( i \right)$ is filled up. Then we finish the construction of $DP\left( i \right)$.
Now, we introduce the detailed processes of constructing an alternative pattern database in \textbf{Algorithm 2}.
\begin{algorithm}
\caption{Constructing an alternative pattern database for the $2$LQR code}
\KwIn{${{S}_{i}}$ ($i=1,\ldots ,{{L}_{b}}$);${{\theta }_{1}}$;${{\theta }_{2}}$}
\KwOut{alternative pattern database}
  \For{$i=1;i\le {{L}_{b}}$}
  {
    Binarize ${{S}_{i}}$ through ${{g}_{1}}\left( {{S}_{i}},{{\theta }_{1}} \right)$ to obtain a binary pattern $B{{S}_{i}}$ ;\\
    Construct a sub-database of the alternative pattern database $DP\left( i \right)$ through ${{g}_{2}}\left( B{{S}_{i}},{{\theta }_{2}} \right)$.
  }
  return alternative pattern database;
\end{algorithm}

\subsection{Our IC Attack for the LCAC $2$D Barcode}
Based on the PDBD scheme described in Section IV, we further propose an advanced IC attack for the LCAC $2$D barcode.
Unlike the detection problem in \eqref{equ5}, our IC attack should pinpoint the exact locations on the received $2$D barcode that are embedded with hidden information.
Thus, we perform a threshold test using the following hypotheses for detecting the presence of the hidden information on the current module, given as
\begin{eqnarray}
\begin{matrix}
   {{\mathcal{H}}_{0}}: & \text{Current module without containing }{{I}_{h}}  \\
   {{\mathcal{H}}_{1}}: & \text{Current module with containing }{{I}_{h}}  \\
\end{matrix}
\label{equ30}
\end{eqnarray}
Specifically, we construct a test statistic for the $j\text{th}$ module ($j=1,\ldots ,{{L}_{d}}$) as
\begin{equation}
\delta \left( j \right)={{\bar{I}}_{o}}\left( j \right)-{{\bar{Y}}_{\text{1}}}\left( j \right),
\label{equ31}
\end{equation}
where ${{L}_{d}}$ is the number of modules in the received $2$D barcode.
We assume that each module consists of ${{N}_{p}}\times {{N}_{p}}$ pixels in the LCAC $2$D barcode.
Here, ${{\bar{I}}_{o}}\left( j \right)$ and ${{\bar{Y}}_{\text{1}}}\left( j \right)$ are the average gray values of the $j\text{th}$ module in ${{\hat{I}}_{o}}$ and ${{\hat{Y}}_{\text{1}}}$, respectively.
By considering an SPC process, it is reasonable to assume ${{\hat{I}}_{o}}={{I}_{o}}$.
%
%
Then, we make a test decision ${{\varphi }_{a}}\left( j \right)$ on $\delta $ according to
\begin{equation}
{{\varphi }_{a}}\left( j \right)=\left\{ \begin{matrix}
   0, & \delta \in {{\theta }_{a}}  \\
   1, & \delta \notin {{\theta }_{a}}  \\
\end{matrix} \right.,
\label{equ34}
\end{equation}
where ${{\theta }_{a}}$ is the range of the test threshold using the hypotheses defined in \eqref{equ30}.
Here, ${{\varphi }_{a}}\left( j \right)=1$ represents the hypothesis ${{\mathcal{H}}_{1}}$ in \eqref{equ30} while ${{\varphi }_{a}}\left( j \right)=0$ represents the opposite hypothesis.
\begin{figure}[!t]
\centering
\includegraphics[width=3.2in]{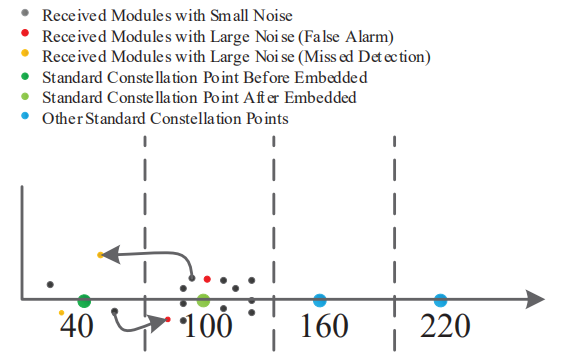}
\caption{Diagram of our IC attack for the LCAC $2$D barcode with 4-order modulation.}
\label{Fig Diagram_LCAC_modulation}
\vspace{-0.5cm}
\end{figure}

The basic idea of our IC attack is derived from a fundamental fact, \ie, provided that the channel noise is small, the received modules should appear around the constellation point, as illustrated in Fig. \ref{Fig Diagram_LCAC_modulation}, where 4-order modulation is considered, \ie, $Q=4$.
In Fig. \ref{Fig Diagram_LCAC_modulation}, the dashed lines represent the decision boundary of demodulation in a multilevel $2$D barcode.
We take one constellation point as an example to illustrate the idea behind our IC attack for the LCAC $2$D barcode, e.g., ‘$40$’ is a standard constellation point in ${{I}_{o}}$ and it is changed to ‘$100$’ in $I$.
If the channel noise is large, there are two types of errors to occur.
The first type is the false alarm, as illustrated by the red dots in Fig. \ref{Fig Diagram_LCAC_modulation}, where the modules belonging to the constellation point of $40$ is mistakenly classified to the constellation point of $100$. The second type is the missed detection, as illustrated by the yellow dots in Fig. \ref{Fig Diagram_LCAC_modulation}, where the modules belonging to the constellation point of $100$ are mistakenly classified to the constellation point of $40$.
Thus, the optimal value of ${{\theta }_{a}}$ in \eqref{equ34} is also determined by making the PFA less than a predetermined upper bound, ${{\varepsilon }_{a,\text{PFA}}}$, according to the NP theorem.

If we can correctly find most of the embedded locations, we can launch an effective IC attack.
Specifically, our IC attack first recovers the original message ${{I}_{o}}$.
Second, our IC attack directly demodulates the received modules at the embedded locations (${{\varphi }_{a}}\left( j \right)=1$) and then replaces the modules of ${{I}_{o}}$ at corresponding locations (${{\varphi }_{a}}\left( j \right)=1$) by the demodulated modules.
At last, our IC attack prints it to launch an effective IC attack. Moreover, according to the results of \cite{Xie2020Low,Villn2006Multilevel}, we know the fact that the channel noises around different constellation points have different distributions.
Thus, we propose two advanced IC attacks for the LCAC $2$D barcode under two cases: All Constellation Points (ACP) and Separate Constellation Points (SCP).
In the ACP case, the detection on embedded locations is designed on all constellation points, where we treat all constellation points equally.
In the SCP case, the detection on embedded locations is separately designed on different constellation points, where we treat different constellation points individually.
Similar to the previous section, we also use GGD RVs to describe the behavior of the test statistics of our IC attacks for obtaining their optimal test thresholds, which are introduced as follows, respectively.

\subsubsection{ACP Case}
\begin{figure}[!t]
\centering
\includegraphics[width=3.55in]{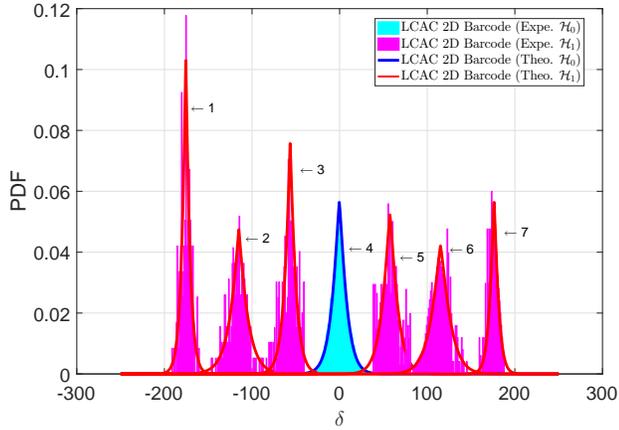}
\caption{Comparison of histograms and theoretical results about the test statistic of our IC attack with the ACP case for the LCAC $2$D barcode, where a camera phone is used as the capturing device.}
\label{Fig LCAC_ACP}
\vspace{-0.5cm}
\end{figure}

We present the distribution of the test statistic of our IC attack with the ACP case for the LCAC $2$D barcode, as illustrated in Fig. \ref{Fig LCAC_ACP}, where both histograms and theoretical results are presented.
The experimental settings and the parameters of the LCAC $2$D barcode are the same as those of Fig. \ref{Fig PDBD} From Fig. \ref{Fig LCAC_ACP}, we observe that the experimental results match well with seven GGD approximations.
We index all GGD RVs as $i=1,\cdots ,7$ from left to right, where the $4$th GGD RV describes the behavior of $\delta $ under ${{\mathcal{H}}_{0}}$ and the remaining GGD RVs together describe the behavior of $\delta $ under ${{\mathcal{H}}_{1}}$.
In comparison with the number of standard constellation points, the number of GGD RVs under ${{\mathcal{H}}_{1}}$  is large because the gray value of each constellation point after embedded will generate three different gray values.
We take the constellation of 40 as an example. When it is embedded by hidden information, its gray value may become to $100$, $160$, or $220$. Correspondingly, the value of $\delta $ in \eqref{equ31} has three different values: -$60$, -$120$, and -$180$, respectively.

Similar to \eqref{equ19}-\eqref{equ23}, we obtain the parameters of each GGD RV, \ie, the mean ${{\mu }_{i}}$, variance $\sigma _{i}^{2}$, and shape factor ${{\gamma }_{{{i}_{}}}}$.
Fig. \ref{Fig LCAC_ACP} shows that the PFA is generated under two situations: left and right. Under the left situation, when we observe the distribution from the left of the $4$th GGD RV to the right, we can calculate the PFA as
\begin{equation}
{{P}_{\text{FA }\!\!\_\!\!\text{ }l}}=\Pr \left\{ {{\left. \delta  \right|}_{{{\mathcal{H}}_{0}}}}<{{\theta }_{a(all\_l)}} \right\}={{F}_{X}}({{\theta }_{a(all\_l)}}),
\label{equ35}
\end{equation}
where ${{\theta }_{a(all\_l)}}$ is the left threshold. Similarly, under the right situation, when we observe the distribution from the right of the $4$th GGD RV to the left, we can calculate the PFA as
\begin{equation}
{{P}_{\text{FA }\!\!\_\!\!\text{ r}}}=\Pr \left\{ {{\left. \delta  \right|}_{{{\mathcal{H}}_{0}}}}>{{\theta }_{a(all\_r)}} \right\}=1-{{F}_{X}}({{\theta }_{a(all\_r)}}),
\label{equ36}
\end{equation}
where ${{\theta }_{a(all\_r)}}$ is the right threshold.
According to the NP theorem, we have ${{P}_{\text{FA }\!\!\_\!\!\text{ }l}}+{{P}_{\text{FA }\!\!\_\!\!\text{ r}}}\le {{\varepsilon }_{a,\text{PFA}}}$.
For convenience, we assume that the PFAs under two situations are similar and both are less than $\frac{{{\varepsilon }_{a,\text{PFA}}}}{2}$, \ie, ${{P}_{\text{FA }\!\!\_\!\!\text{ }l}}\le \frac{{{\varepsilon }_{a,\text{PFA}}}}{2}$ and ${{P}_{\text{FA }\!\!\_\!\!\text{ r}}}\le \frac{{{\varepsilon }_{a,\text{PFA}}}}{2}$.
Then, we respectively obtain the optimal value of ${{\theta }_{a(all\_l)}}$ and ${{\theta }_{a(all\_r)}}$ as
\begin{equation}
\theta _{a(all\_l)}^{0}=-\frac{{{\kappa }^{-1}}\left[ \frac{\text{1}}{{{\gamma }_{4}}},\left( 1-{{\varepsilon }_{a,\text{PFA}}} \right)\Gamma \left( {1}/{{{\gamma }_{4}}}\; \right) \right]}{\eta ({{\sigma }_{4}},{{\gamma }_{4}}){{\gamma }_{4}}}\text{+}{{\mu }_{4}},
\label{equ37}
\end{equation}
\begin{equation}
\theta _{a(all\_r)}^{0}=\frac{{{\kappa }^{-1}}\left[ \frac{\text{1}}{{{\gamma }_{4}}},\left( 1-{{\varepsilon }_{\text{PFA}}} \right)\Gamma \left( {1}/{{{\gamma }_{4}}}\; \right) \right]}{\eta ({{\sigma }_{4}},{{\gamma }_{4}}){{\gamma }_{4}}}\text{+}{{\mu }_{4}}.
\label{equ38}
\end{equation}
In summary, the range of the optimal threshold in our IC attack with the ACP case is denoted as
\begin{equation}
\theta _{a(all)}^{0}=\left[ \theta _{a(all\_l)}^{0},\theta _{a(all\_r)}^{0} \right].
\label{equ39}
\end{equation}
When the value of $\delta $ falls into $\theta _{a(all)}^{0}$, we accept ${{\mathcal{H}}_{0}}$ in \eqref{equ30} and we set ${{\varphi }_{a}}\left( j \right)=0$;
otherwise, we accept ${{\mathcal{H}}_{1}}$ in \eqref{equ30} and we set ${{\varphi }_{a}}\left( j \right)=1$.

\subsubsection{SCP Case}
\begin{figure}[!t]
\centering
\includegraphics[width=3.75in]{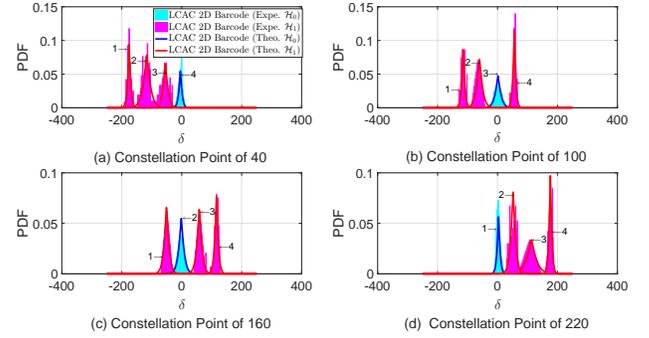}
\caption{Comparison of histograms and theoretical results about the test statistic of our IC attack with the SCP case for the LCAC $2$D barcode under four scenarios: (a) Constellation Point of $40$; (b) Constellation Point of $100$; (c) Constellation Point of $160$; (4) Constellation Point of $220$, where a camera phone is used as the capturing device.}
\label{Fig LCAC_SCP}
\vspace{-0.5cm}
\end{figure}

Since the channel noises around different constellation points have different distributions, we present the distribution of the test statistic of our IC attack with the SCP case for the LCAC $2$D barcode, as illustrated in Fig. \ref{Fig LCAC_SCP}, where both histograms and theoretical results are presented.
The experimental settings and the parameters of the LCAC $2$D barcode are the same as those of Fig. \ref{Fig PDBD}.
From each subfigure of Fig. \ref{Fig LCAC_SCP}, we observe that the experimental results match well with four GGD approximations.
We index all GGD RVs as $i=1,\cdots ,4$ from left to right.
Then, similar to \eqref{equ19}-\eqref{equ23}, we obtain the parameters of each GGD RV, \ie, the mean ${{\mu }_{i}}$, variance $\sigma _{i}^{2}$, and shape factor ${{\gamma }_{{{i}_{}}}}$.
Based on the distribution of different scenarios illustrated in each subfigure of Fig. \ref{Fig LCAC_SCP}, we perform a threshold test to detect the presence of the hidden information on the current module under different scenarios.

Similar to our IC attack with the ACP case, we derive the optimal value of ${{\theta }_{a}}$ for our IC attack with the SCP case under different scenarios by making ${{P}_{\text{FA}}}\le {{\varepsilon }_{a,\text{PFA}}}$.
Specifically, for the scenario of the constellation point of $40$, as illustrated in Fig. \ref{Fig LCAC_SCP}(a), we obtain the range of the optimal threshold as
\begin{equation}
\theta _{a(40)}^{0}\ge -\frac{{{\kappa }^{-1}}\left[ \frac{\text{1}}{{{\gamma }_{4}}},\left( 1-2{{\varepsilon }_{a,\text{PFA}}} \right)\Gamma \left( {1}/{{{\gamma }_{4}}}\; \right) \right]}{\eta ({{\sigma }_{4}},{{\gamma }_{4}}){{\gamma }_{4}}}\text{+}{{\mu }_{4}}.
\label{equ40}
\end{equation}
For the scenario of the constellation point of $100$, as illustrated in Fig. \ref{Fig LCAC_SCP}(b), we obtain the optimal range of ${{\theta }_{a(100\_l)}}$ and ${{\theta }_{a(100\_r)}}$ as
\begin{equation}
\theta _{a(100\_l)}^{0}=-\frac{{{\kappa }^{-1}}\left[ \frac{\text{1}}{{{\gamma }_{3}}},\left( 1-{{\varepsilon }_{a,\text{PFA}}} \right)\Gamma \left( {1}/{{{\gamma }_{3}}}\; \right) \right]}{\eta ({{\sigma }_{3}},{{\gamma }_{3}}){{\gamma }_{3}}}\text{+}{{\mu }_{3}},
\label{equ41}
\end{equation}
\begin{equation}
\theta _{a(100\_r)}^{0}=\frac{{{\kappa }^{-1}}\left[ \frac{\text{1}}{{{\gamma }_{3}}},\left( 1-{{\varepsilon }_{a,\text{PFA}}} \right)\Gamma \left( {1}/{{{\gamma }_{3}}}\; \right) \right]}{\eta ({{\sigma }_{3}},{{\gamma }_{3}}){{\gamma }_{3}}}\text{+}{{\mu }_{3}}.
\label{equ42}
\end{equation}
Then, we obtain the range of the optimal threshold as
\begin{equation}
\theta _{a(100)}^{0}=\left[ \theta _{a(100\_l)}^{0},\theta _{a(100\_r)}^{0} \right].
\label{equ43}
\end{equation}
For the scenario of the constellation point of $160$, as illustrated in Fig. \ref{Fig LCAC_SCP}(c), we obtain the optimal range of ${{\theta }_{a(160\_l)}}$ and ${{\theta }_{a(160\_r)}}$ as
\begin{equation}
\theta _{a(160\_l)}^{0}=-\frac{{{\kappa }^{-1}}\left[ \frac{\text{1}}{{{\gamma }_{2}}},\left( 1-{{\varepsilon }_{a,\text{PFA}}} \right)\Gamma \left( {1}/{{{\gamma }_{2}}}\; \right) \right]}{\eta ({{\sigma }_{2}},{{\gamma }_{2}}){{\gamma }_{2}}}\text{+}{{\mu }_{2}},
\label{equ44}
\end{equation}
\begin{equation}
\theta _{a(160\_r)}^{0}=\frac{{{\kappa }^{-1}}\left[ \frac{\text{1}}{{{\gamma }_{2}}},\left( 1-{{\varepsilon }_{a,\text{PFA}}} \right)\Gamma \left( {1}/{{{\gamma }_{2}}}\; \right) \right]}{\eta ({{\sigma }_{2}},{{\gamma }_{2}}){{\gamma }_{2}}}\text{+}{{\mu }_{2}}.
\label{equ45}
\end{equation}
Then, we obtain the range of the optimal threshold as
\begin{equation}
\theta _{a(160)}^{0}=\left[ \theta _{a(160\_l)}^{0},\theta _{a(160\_r)}^{0} \right].
\label{equ46}
\end{equation}
For the scenario of the constellation point of $220$, as illustrated in Fig. \ref{Fig LCAC_SCP}(d), we obtain the range of the optimal threshold as
\begin{equation}
\theta _{a\left( 220 \right)}^{0}\le \frac{{{\kappa }^{-1}}\left[ \frac{\text{1}}{{{\gamma }_{1}}},\left( 1-2{{\varepsilon }_{a,\text{PFA}}} \right)\Gamma \left( {1}/{{{\gamma }_{1}}}\; \right) \right]}{\eta ({{\sigma }_{1}},{{\gamma }_{1}}){{\gamma }_{1}}}\text{+}{{\mu }_{1}}.
\label{equ47}
\end{equation}
When the value of $\delta $ falls into $\theta _{a(40)}^{0}$, or $\theta _{a\left( 100 \right)}^{0}$, or $\theta _{a\left( 160 \right)}^{0}$, or $\theta _{a\left( 220 \right)}^{0}$, we accept ${{\mathcal{H}}_{0}}$ in \eqref{equ30} and we set ${{\varphi }_{a}}\left( j \right)=0$;
otherwise, we accept ${{\mathcal{H}}_{1}}$ in \eqref{equ30} and we set ${{\varphi }_{a}}\left( j \right)=1$.

%
\section{Experiment Results of Advanced IC Attacks}

\subsection{Experimental Setup}
\begin{figure}[!t]
\centering
\includegraphics[width=3.3in]{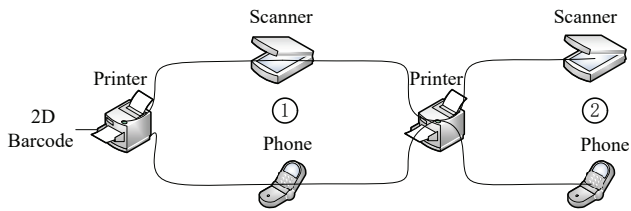}
\caption{Equipment settings for emulating a DPC process.}
\label{Fig Equipment_setting}
\vspace{-0.5cm}
\end{figure}

Note that in both Section IV and Section V, we consider an SPC process in the experimental results.
However, since this section emulates IC attacks for two existing anti-copying $2$D barcodes, we should consider a DPC process in the experimental results.
We illustrate the equipment setting for emulating a DPC process in Fig. \ref{Fig Equipment_setting}. Specifically, first, a sender prints a legitimate $2$D barcode according to the principle defined in existing anti-copying $2$D barcodes.
Second, a monitor captures it by using the first capturing device, \eg, a scanner or a camera phone, and then prints an illegitimate $2$D barcode by launching our advanced IC attacks.
At last, a receiver captures the illegitimate $2$D barcode by using the second capturing device, \eg, a scanner or a camera phone, and then makes an authentication test defined in \eqref{equ4}.

We consider two existing anti-copying $2$D barcodes:
\begin{enumerate}
  \item $2$LQR code, where we set the authentication threshold of the Pearson correlation as ${{\theta }_{b}}=0.12$, just like in \cite{Tkachenko2016Two};
  \item LCAC $2$D barcode, where we set the error correction capability of authentication message as ${{t}_{a}}=10$ and we set the authentication threshold of the BER as ${{\theta }_{b}}=1.2\%$, just like in \cite{Xie2020Low}.
\end{enumerate}
Moreover, we consider five IC attacks:
\begin{enumerate}
  \item Direct IC attack;
  \item Synthetic IC attack, where 6 versions of a legitimate $2$D barcode are used to synthesize an illegitimate $2$D barcode;
  \item ML-based IC attack \cite{Yadav2019Estimation}, where for the $2$LQR code, we use the equalization outputs of ${{L}_{m}}$ leg3)	ML-based IC attack \cite{Yadav2019Estimation}, where for the $2$LQR code, we use the equalization outputs of ${{L}_{m}}$ legitimate $2$D barcodes, \ie, ${{\hat{Y}}_{1}}$, and the electronic version of ${{L}_{m}}$ corresponding legitimate $2$D barcode, \ie, $I$, to obtain an attacking model;
      Similarly, for the LCAC $2$D barcode, we use the equalization outputs of $60$ legitimate $2$D barcodes, and the electronic version of $60$ corresponding legitimate $2$D barcodes to obtain an attacking model.
  \item Our IC attacks for the $2$LQR code under the PPD and UPD scenarios, where we set ${{\theta }_{1}}=127$ and ${{\theta }_{2}}\text{=}0.8$ for the UPD scenario;
  \item Our IC attack for the LCAC $2$D barcode, where we set ${{\varepsilon }_{a,\text{PFA}}}=0.01$.
\end{enumerate}
The experimental settings and the parameters of two existing anti-copying $2$D barcodes are the same as those given in Section IV.C.
Here we consider two performance metrics:
\begin{enumerate}
  \item The first metric is the probability of successfully attacking, which is defined as
\begin{equation}
{{P}_{s}}\text{=}\frac{{{M}_{a}}}{{{M}_{r}}},
\label{equ48}
\end{equation}
where ${{M}_{r}}$ is the number of received illegitimate $2$D barcodes and ${{M}_{a}}$ is the number of illegitimate $2$D barcodes mistakenly accepted by the receiver as legitimate ones;
  \item The second metric is the probability of correctly pinpointing the embedded locations in the LCAC $2$D barcode, which is defined as
\begin{equation}
{{P}_{r}}\text{=}\frac{{{L}_{c}}}{{{L}_{h}}}.
\label{equ49}
\end{equation}
We define ${{\mathcal{C}}_{h}}$ as the set of actual embedded locations in hidden information and ${{L}_{h}}$ is the corresponding set length.
We define ${{\mathcal{C}}_{e}}$ as the set of estimated embedded locations, ${{\mathcal{C}}_{c}}$ is the intersection set of ${{\mathcal{C}}_{e}}$ and ${{\mathcal{C}}_{h}}$, \ie, ${{\mathcal{C}}_{c}}={{\mathcal{C}}_{e}}\bigcap {{\mathcal{C}}_{h}}$, ${{L}_{c}}$ is the set length of ${{\mathcal{C}}_{c}}$.
\end{enumerate}
\subsection{Experimental Results for the $2$LQR Code}

We present the experimental results for the $2$LQR Code under the PPD and UPD scenarios, respectively, where we use a scanner as both the first and second capturing devices.
\subsubsection{PPD Scenario}
\begin{table}
\centering
\caption{Comparison of ${{P}_{s}}$ of various IC attacks for the $2$LQR code under the PPD scenario.}
\footnotesize
\begin{tabular}{|c|c|c|c|c|}
  \hline
  IC Attacks & Direct & Synthetic & ML-Based & Our \\
  \hline
  ${{P}_{s}}$ & $0$ & $0$ & $0.40$ &  $0.6538$ \\
  \hline
\end{tabular}
\label{Table IC PPD}
\vspace{0.2cm}
\end{table}

We first compare the values of ${{P}_{s}}$ of various IC attacks for the $2$LQR code under the PPD scenario, as presented in Tab. \ref{Table IC PPD}, where we set ${{L}_{m}}\text{=}100$. From Tab. \ref{Table IC PPD}, we observe that our IC attack has the best performance but both the direct and synthetic IC attacks have the poorest performance.
Although the ML-based IC attack requires a training stage, its performance is inferior to our IC attack since the output patterns of the ML-based IC attack may not be exactly the same as those used in the $2$LQR code.
\begin{table}
\centering
\caption{Comparison of ${{P}_{s}}$ of our IC attack versus different ${{L}_{m}}$ for the $2$LQR code under the PPD scenario.}
\footnotesize
\begin{tabular}{|c|c|c|c|}
  \hline
  ${{L}_{m}}$ & $100$ & $1000$ & $10000$  \\
  \hline
  ${{P}_{s}}$ & $0.6538$ & $0.6226$ & $0.5637$  \\
  \hline
\end{tabular}
\label{Table PPD}
\vspace{0.2cm}
\end{table}

Now, we investigate the impact of ${{L}_{m}}$ on the performance of our IC attack under the PPD scenario, as presented in Tab. \ref{Table PPD}, where we use the same experimental settings of Tab. \ref{Table IC PPD} except different ${{L}_{m}}$. From Tab. \ref{Table PPD}, we observe that the performance of our IC attack declines as the value of ${{L}_{m}}$ increasing.
This is because, given the size of a pattern, the distance between different patterns reduces as the value of ${{L}_{m}}$ increasing, which lowers the discriminability of the way of comparing the Pearson correlation.
\subsubsection{UPD Scenario}
\begin{table}
\centering
\caption{Comparison of ${{P}_{s}}$ of various IC attacks for the $2$LQR code under the UPD scenario.}
\footnotesize
\begin{tabular}{|c|c|c|c|c|}
  \hline
  IC Attacks & Direct & Synthetic & ML-Based & Our \\
  \hline
  ${{P}_{s}}$ & $0$ & $0$ & $0.281$ &  $0.294$ \\
  \hline
\end{tabular}
\label{Table IC UPD}
\vspace{0.2cm}
\end{table}

We first compare the values of ${{P}_{s}}$ of various IC attacks for the $2$LQR code under the UPD scenario, as presented in Tab. \ref{Table IC UPD}, where we use the same experimental settings of Tab. \ref{Table IC PPD} except that the public pattern database is replaced by an alternative pattern database.
The ML-based IC attack straightforwardly uses the alternative pattern database to train an attacking model.
By comparing the results of Tab. \ref{Table IC PPD} with those of Tab. \ref{Table IC UPD}, we observe that the performances of all IC attacks become worse since the patterns used in the $2$LQR code may not exist in the alternative pattern database.
Moreover, from Tab. \ref{Table IC UPD}, we can obtain the same conclusions of Tab. \ref{Table IC PPD} because of the same reasons.
Specifically, our IC attack has the best performance, the ML-based IC attack is the second, and both the direct and synthetic IC attacks have the poorest performance.
\begin{table}
\centering
\caption{Comparison of ${{P}_{s}}$ of our IC attack versus different ${{L}_{m}}$ for the $2$LQR code under the UPD scenario.}
\footnotesize
\begin{tabular}{|c|c|c|c|}
  \hline
  ${{L}_{m}}$ & $100$ & $1000$ & $10000$  \\
  \hline
  ${{P}_{s}}$ & $0.294$ & $0.309$ & $0.325$  \\
  \hline
\end{tabular}
\label{Table UPD}
\vspace{0.2cm}
\end{table}

Now, we investigate the impact of ${{L}_{m}}$ on the performance of our IC attack under the UPD scenario, as presented in Tab. \ref{Table UPD}, where we use the same experimental settings of Tab. \ref{Table PPD} except that the public pattern database is replaced by an alternative pattern database.
By comparing the results of Tab. \ref{Table PPD} with those of Tab. \ref{Table UPD}, we observe that the performances of all IC attacks become worse because of the same reasons of Tab. \ref{Table IC UPD}.
Moreover, from Tab. \ref{Table UPD}, we observe an opposite conclusion of Tab. \ref{Table PPD}.
Specifically, the performance of our IC attack improves as the value of ${{L}_{m}}$ increasing.
This is because the probability of the patterns used in the $2$LQR code occurring in the alternative pattern database increases as the value of ${{L}_{m}}$ increasing.
\subsection{Experimental Results for the LCAC $2$D Barcode}
\begin{table}
\centering
\caption{Comparison of ${{P}_{r}}$ of our IC attacks for the LCAC $2$D barcode under the ACP and SCP cases.}
\footnotesize
\begin{tabular}{|c|c|c|c|c|}
  \hline
  First Capturing Devices & \multicolumn{2}{c|}{Scanner}  & \multicolumn{2}{c|}{Camera Phone} \\
  \hline
  Cases & ACP & SCP & ACP & SCP \\
  \hline
  ${{P}_{r}}$ & $0.9701$ & $0.9830$ & $0.9725$ &  $0.9852$ \\
  \hline
\end{tabular}
\label{Table ACP and SCP}
\vspace{0.2cm}
\end{table}

We first compare the values of ${{P}_{r}}$ of our IC attacks for the LCAC $2$D barcode under the ACP and SCP cases, as presented in Tab. \ref{Table ACP and SCP}, where we use both a scanner and a camera phone as the first capturing devices and we set ${{\varepsilon }_{a,\text{PFA}}}=0.01$.
From Tab. \ref{Table ACP and SCP}, we observe that our IC attack with the SCP case has better pinpointing accuracy than that of our IC attack with the ACP case.
Moreover, the pinpointing accuracy under the camera phone is better than that under the scanner since the camera phone has better-capturing resolution than the scanner in our experimental setup.
\begin{table}
\centering
\caption{Comparison of ${{P}_{s}}$ of various IC attacks for the LCAC $2$D barcode under the ACP and SCP cases with different ${{t}_{a}}$.}
\footnotesize
\begin{tabular}{|c|c|c|c|c|}
  \hline
  ${{t}_{a}}$ & \multicolumn{2}{c|}{10}  & \multicolumn{2}{c|}{8} \\
  \hline
  \tabincell{c}{Second Capturing\\Device}  & Scanner &\tabincell{c}{Camera\\Phone}& Scanner & \tabincell{c}{Camera\\Phone} \\
  \hline
  Direct & $0.0909$ & $0.1102$ & $0.0872$ &  $0.1023$ \\
  \hline
  Synthetic & $0.8051$ & $0.8549$ & $0.7835$ &  $0.8325$ \\
  \hline
  Our (ACP) & $1$ & $1$ & $0.9732$ &  $0.98$ \\
  \hline
  Our (SCP) & $1$ & $1$ & $1$ &  $1$ \\
  \hline
\end{tabular}
\label{Table IC ACP and SCP}
\vspace{0.2cm}
\end{table}

Now we compare the values of ${{P}_{s}}$ of various IC attacks for the LCAC $2$D barcode under the ACP and SCP cases with different ${{t}_{a}}$, as presented in Tab. \ref{Table IC ACP and SCP}, where we use the same experimental settings of Tab. VII except that we use a scanner as the first capturing devices, and use both a scanner and a camera phone as the second capturing devices.
From Tab. \ref{Table IC ACP and SCP}, we observe the following conclusions.
First, when ${{t}_{a}}=10$, our IC attacks under both the ACP and SCP cases have the best performance, the synthetic IC attack is the second one, and the direct IC attack is the last one.
Second, when the value of ${{t}_{a}}$ reduces, \ie, the error correction capability of authentication message in the LCAC $2$D barcode declines, the performance of all IC attacks declines except our IC attack under the SCP case, which highlights the advantage of our IC attack under the SCP case.
Third, the attacking performance of all IC attacks under the camera phone is better than that under the scanner since the camera phone has better-capturing resolution than the scanner in our experimental setup.
\subsection{Discussion}
\begin{table}
\centering
\caption{Comparison of ${{P}_{s}}$ of various IC attacks for both the $2$LQR code and the LCAC $2$D barcode.}
\footnotesize
\begin{tabular}{|c|c|c|c|c|}
  \hline
  $2$D Barcode & \multicolumn{2}{c|}{$2$LQR Code}  & \multicolumn{2}{c|}{LCAC $2$D Barcode ($Q=2$)} \\
  \hline
  \tabincell{c}{Second Capturing\\Device} & Scanner & \tabincell{c}{Camera\\Phone} & Scanner & \tabincell{c}{Camera\\Phone} \\
  \hline
  Direct  & $0$ & - & $0.9630$ &  $0.9850$ \\
  \hline
  Synthetic  & $0$ & - & $1$ &  $1$ \\
  \hline
  ML Based & $0.40$ & - & $1$ &  $1$ \\
  \hline
  Our & $0.6538$ & - & $1$ &  $1$ \\
  \hline
\end{tabular}
\label{Table IC Q=$2$}
\vspace{0.2cm}
\end{table}

Since the ML-based IC attack \cite{Yadav2019Estimation} was designed for standard $2$D barcode, \ie, $Q=2$, we did not present the corresponding experimental results in Tab. \ref{Table IC ACP and SCP}.
For fairly comparing the performance of various IC attacks on both the $2$LQR code and the LCAC $2$D barcode, we extend the LCAC $2$D barcode to the scenario of $Q=2$, where we set ${{t}_{a}}=1$, just like in \cite{Xie2020Low}.
We present the experimental results in Tab. \ref{Table IC Q=$2$}, where the PPD scenario is considered and ${{L}_{m}}\text{=}100$ for the $2$LQR code, the SCP case is considered for the LCAC $2$D barcode, and the remaining experimental settings are the same as those of Tab. \ref{Table IC ACP and SCP}.
From Tab. \ref{Table IC Q=$2$}, we observe the following conclusions.
First, our IC attack has the same attacking performance as that of the ML-based IC attack for the LCAC $2$D barcode whereas our IC attack has better attacking performance as that of the ML-based IC attack for the $2$LQR code.
Second, the performance of all IC attacks improves at the modulation order reducing for the LCAC $2$D barcode as compared with the results of Tab. \ref{Table IC ACP and SCP}.
Third, our IC attack has better attacking performance for the LCAC $2$D barcode than that for the $2$LQR code, however, the $2$LQR code cannot be applied to the scenario of a camera phone since the $2$LQR code requires higher stability of the capturing equipment.
\section{Conclusion}
In this paper, we proposed two hidden information detection schemes at the existing anti-copying $2$D barcodes.
We explicitly analyzed the PD and PFA of our schemes, derived their closed-form expressions, and obtained their optimal test thresholds.
Besides we provided the experimental results of our detection schemes under different capturing devices, we verified the above theoretical results with simulation results.
Moreover, we designed advanced IC attacks to evaluate the security of two existing anti-copying $2$D barcodes.
Note that our IC attacks can deal with the scenario of an unknown pattern database in the $2$LQR code and can correctly pinpoint the embedded locations in the LCAC $2$D barcode.
We implemented our IC attacks and conducted extensive performance comparison between our schemes and prior schemes under different capturing devices.
Interesting future research is to detect the existence of the hidden information in a printed $2$D barcode when a monitor does not know the parameters of modulation and channel coding of the original message, which becomes more challenging.

\bibliographystyle{IEEEtran}
\bibliography{IEEEabrv,MyRef}

\begin{thebibliography}{10}
\providecommand{\url}[1]{#1}
\csname url@samestyle\endcsname
\providecommand{\newblock}{\relax}
\providecommand{\bibinfo}[2]{#2}
\providecommand{\BIBentrySTDinterwordspacing}{\spaceskip=0pt\relax}
\providecommand{\BIBentryALTinterwordstretchfactor}{4}
\providecommand{\BIBentryALTinterwordspacing}{\spaceskip=\fontdimen2\font plus
\BIBentryALTinterwordstretchfactor\fontdimen3\font minus
  \fontdimen4\font\relax}
\providecommand{\BIBforeignlanguage}[2]{{%
\expandafter\ifx\csname l@#1\endcsname\relax
\typeout{** WARNING: IEEEtran.bst: No hyphenation pattern has been}%
\typeout{** loaded for the language `#1'. Using the pattern for}%
\typeout{** the default language instead.}%
\else
\language=\csname l@#1\endcsname
\fi
#2}}
\providecommand{\BIBdecl}{\relax}
\BIBdecl

\bibitem{Tkachenko2016Two}
I.~Tkachenko, W.~Puech, C.~Destruel, O.~Strauss, J.~Gaudin, and C.~Guichard,
  ``Two-level qr code for private message sharing and document
  authentication,'' \emph{IEEE Transactions on Information Forensics \&
  Security}, vol.~11, no.~3, pp. 571--583, 2016.

\bibitem{Xie2020Low}
N.~Xie, Q.~Zhang, J.~Hu, G.~Luo, and C.~Chen, ``Low-cost anti-copying 2d
  barcode by exploiting channel noise characteristics,'' \emph{arXiv e-prints},
  Jan. 2020.

\bibitem{Yadav2019Estimation}
R.~Yadav, T.~Iuliia, A.~Treméau, and T.~Fournel, ``Estimation of
  copy-sensitive codes using a neural approach,'' in \emph{ACM Workshop on
  Information Hiding and Multimedia Security (IH\&MMSec)}, Paris, France, 2019,
  pp. 77--82.

\bibitem{Adams2011A}
G.~B. Adams, S.~B. Pollard, and S.~J. Simske, ``A study of the interaction of
  paper substrates on printed forensic imaging,'' in \emph{ACM Symposium on
  Document Engineering (DocEng)}, Mountain View, California, USA, 2011, pp.
  36--40.

\bibitem{Simske2010High}
S.~J. Simske and G.~B. Adams, ``High-resolution glyph-inspection based security
  system,'' in \emph{IEEE International Conference on Acoustics, Speech and
  Signal (ICASSP)}, Dallas, Texas, USA, 2010, pp. 1794--1797.

\bibitem{Gremaud2015Identification}
X.~Marguerettaz, F.~Gremaud, A.~Commeureuc, V.~Aboutanos, T.~Tiller, and
  O.~Rozumek, ``Identification and authentication using liquid crystal material
  markings,'' Jun.~3 2014, uS Patent 8,740,088.

\bibitem{Wang2013Using}
H.~C. Wang, Y.~W. Cheng, W.~C. Huang, C.~L. Chang, and S.~Y. Lu, ``Using
  modified digital halftoning technique to design invisible 2d barcode by
  infrared detection,'' \emph{Lecture Notes in Electrical Engineering}, vol.
  234, no.~2, pp. 179--186, 2013.

\bibitem{Voloshynovskiy2016Physical}
S.~Voloshynovskiy, T.~Holotyak, and P.~Bas, ``Physical object authentication:
  detection-theoretic comparison of natural and artificial randomness,'' in
  \emph{IEEE International Conference on Acoustics, Speech and Signal
  Processing (ICASSP)}, Shanghai, China, 2016, pp. 2029--2033.

\bibitem{Voloshynovskiy2012Towards}
S.~Voloshynovskiy, M.~Diephuis, F.~Beekhof, and O.~Koval, ``Towards
  reproducible results in authentication based on physical non-cloneable
  functions: The forensic authentication microstructure optical set
  ({FAMOS)},'' in \emph{IEEE International Workshop on Information Forensics
  and Security (WIFS)}, Tenerife, Spain, 2012, pp. 43--48.

\bibitem{Wong2016Counterfeit}
C.~W. Wong and M.~Wu, ``Counterfeit detection using paper {PUF} and mobile
  cameras,'' in \emph{IEEE International Workshop on Information Forensics and
  Security (WIFS)}, Rome, Italy, 2016, pp. 1--6.

\bibitem{Wong2015A}
------, ``A study on {PUF} characteristics for counterfeit detection,'' in
  \emph{IEEE International Conference on Image Processing (ICIP)}, Quebec City,
  Quebec, Canada, 2015, pp. 1643--1647.

\bibitem{Diephuis2013Physical}
M.~Diephuis and S.~Voloshynovskiy, ``Physical object identification based on
  {FAMOS} microstructure fingerprinting: comparison of templates versus
  invariant features,'' in \emph{International Symposium on Image and Signal
  Processing and Analysis (ISPA)}, Trieste, Italy, 2013, pp. 119--123.

\bibitem{Claycomb2006Using}
W.~Claycomb and D.~Shin, ``Using a two dimensional colorized barcode solution
  for authentication in pervasive computing,'' in \emph{ACS/IEEE International
  Conference on Pervasive Services (ICPS)}, lyon, Rhone-Alpes, France, 2006,
  pp. 173--180.

\bibitem{Picard2004Society}
J.~Picard, ``Digital authentication with copydetection patterns,'' in
  \emph{Society of Photo-Optical Instrumentation Engineers (SPIE)}, San Jose,
  California, USA, 2004, pp. 176--183.

\bibitem{Goldman1987Non}
R.~N. Goldman, ``Non-counterfeitable document system,'' 1987, uS Patent
  4,423,415.

\bibitem{Jorge2019A}
C.-Z. Jorge and G.~Antonio-Javier, ``A selectional auto-encoder approach for
  document image binarization,'' \emph{Pattern Recognition}, vol.~86, no.~1,
  pp. 37--47, 2019.

\bibitem{Zhang2019Accurate}
L.~Zhang, C.~Chen, and W.~H. Mow, ``Accurate modeling and efficient estimation
  of the print-capture channel with application in barcoding,'' \emph{IEEE
  Transactions on Image Processing}, vol.~28, no.~1, pp. 464--478, 2019.

\bibitem{Younan2010Moments}
W.~Y. Younan and K.~S. Selim, ``Moments of order statistics of a generalized
  normal distribution,'' \emph{Far East Journal of Theoretical Statistics},
  vol.~33, no.~1, pp. 93--106, 2010.

\bibitem{Wang2006A}
T.~Wang, H.~Li, Z.~Li, and Z.~Wang, ``A fast parameter estimation of
  generalized gaussian distribution,'' in \emph{international Conference on
  Signal Processing (ICSP)}, vol.~1, Beijing, China, 2006.

\bibitem{Baierlein2004Probability}
R.~Baierlein, ``Probability theory: the logic of science,'' \emph{Mathematical
  Intelligencer}, vol.~57, no.~10, pp. 76--77, 2004.

\bibitem{Sharifi1995Estimation}
K.~Sharifi and A.~Leongarcia, ``Estimation of shape parameter for generalized
  gaussian distributions in subband decompositions of video,'' \emph{IEEE
  Transactions on Circuits \& Systems for Video Technology}, vol.~5, no.~1, pp.
  52--56, 1995.

\bibitem{Villn2006Multilevel}
R.~Villn, S.~Voloshynovskiy, O.~Koval, and T.~Pun, ``Multilevel 2d barcodes:
  toward high-capacity storage modules for multimedia security and
  management,'' \emph{IEEE Transactions on Information Forensics and Security},
  vol.~1, no.~4, pp. 405--420, 2006.

\end{thebibliography}

\end{document}